\documentclass{elsart4-1}

\usepackage{amssymb}
\usepackage{amsfonts}
\usepackage{amsmath}

\usepackage[english,francais]{babel}

\newtheorem{e-proposition}[theorem]{Proposition}

\newtheorem{e-definition}[theorem]{Definition\rm}

\renewcommand{\theequation}{\arabic{equation}}
\def\og{\leavevmode\raise.3ex\hbox{$\scriptscriptstyle\langle\!\langle$~}}
\def\fg{\leavevmode\raise.3ex\hbox{~$\!\scriptscriptstyle\,\rangle\!\rangle$}}

\begin{document}

\centerline{Physics or Astrophysics/Header}
\begin{frontmatter}


\selectlanguage{english} \title{General Relativistic Dynamics of Compact
  Binary Systems}


\selectlanguage{english}
\author[authorlabel1]{Luc Blanchet},
\ead{blanchet@iap.fr}

\address[authorlabel1]{${\mathcal{G}}{\mathbb{R}}
\varepsilon{\mathbb{C}}{\mathcal{O}}$, Institut d'Astrophysique de
Paris, \\ UMR 7095 CNRS, Universit\'e Pierre \& Marie Curie, \\ 98 bis
boulevard Arago, 75014 Paris, France}


\medskip
\begin{center}
\end{center}

\begin{abstract}
The equations of motion of compact binary systems have been derived in
the post-Newtonian (PN) approximation of general relativity. The current
level of accuracy is 3.5PN order. The conservative part of the equations
of motion (neglecting the radiation reaction damping terms) is deducible
from a generalized Lagrangian in harmonic coordinates, or equivalently
from an ordinary Hamiltonian in ADM coordinates. As an application we
investigate the problem of the dynamical stability of circular binary
orbits against gravitational perturbations up to the 3PN order. We find
that there is no innermost stable circular orbit or ISCO at the 3PN
order for equal masses.


\vskip 0.5\baselineskip

\selectlanguage{francais}
\noindent{\bf R\'esum\'e}
\vskip 0.5\baselineskip
\noindent
Les \'equations du mouvement d'un syst\`eme binaire d'objets compacts
ont \'et\'ees calcul\'ees dans l'approximation post-newtonienne (PN) de
la relativit\'e g\'en\'erale. Le niveau d'approximation atteint l'ordre
3.5PN. La partie conservative des \'equations du mouvement (obtenue en
n\'egligeant les termes de freinage de rayonnement) se d\'eduit d'un
lagrangien g\'en\'eralis\'e en coordonn\'ees harmoniques, et, de
fa\c{c}on \'equivalente, d'un hamiltonien ordinaire en coordonn\'ees
ADM. Comme application nous \'etudions le probl\`eme de la stabilit\'e
dynamique, vis-\`a-vis des perturbations gravitationnelles, des orbites
binaires circulaires \`a l'ordre 3PN. Nous trouvons qu'il n'y a pas de
derni\`ere orbite stable circulaire ou ISCO \`a l'ordre 3PN pour des
masses \'egales.


 \vskip 0.5\baselineskip
\noindent{\small{\it Mots-cl\'es~:} Th\'eorie post-newtonienne;
\'equations du mouvement; syst\`emes binaires compacts}}
\end{abstract}
\end{frontmatter}


\selectlanguage{english}

\section{Introduction}\label{secI}

The problem of the dynamics of two compact bodies is part of a program
aimed at unravelling the information contained in the
gravitational-wave signals emitted by inspiralling and coalescing
compact binaries --- systems of neutron stars or black holes driven
into coalescence by emission of gravitational radiation. The treatment
of the problem is Post-Newtonian, \textit{i.e.} based on formal
expansions, when the speed of light $c$ tends to infinity, of general
relativity theory around Newton's theory. The early, classic works of
Lorentz \& Droste~\cite{LD17}, Eddington \& Clark~\cite{EC38},
Einstein, Infeld \& Hoffmann~\cite{EIH}, Fock~\cite{Fock},
Papapetrou~\cite{Papa51} and others led to a good understanding of the
equations of motion of $N$ bodies at the first post-Newtonian
approximation (1PN, corresponding to order~$\sim 1/c^2$). In the
1970's, a series of works~\cite{OO73,OO74a,OO74b} led to a nearly
complete control of the problem of motion at the second post-Newtonian
approximation (2PN~$\sim 1/c^4$). Then, starting in the early 80's,
motivated by the observation of secular orbital effects in the
Hulse-Taylor binary pulsar PSR1913+16~\cite{TFMc79,T93}, several
groups solved the two-body problem at the 2.5PN level (while
completing on the way the derivation of the 2PN
dynamics)~\cite{BeDD81,DD81a,DD81b,D83houches,DS85,GKop86,BFP98,IFA01}. The
2.5PN term constitutes the first contribution of gravitational
\textit{radiation reaction} in the equations of motion (such term is
the analogue of the Abraham--Lorentz reaction force in
electromagnetism), and is directly responsible for the decrease of the
binary pulsar orbital period by emission of gravitational radiation.

In the late 90's, motivated by the aim of deriving high-accuracy
templates for the data analysis of the international network of
interferometric gravitational-wave detectors LIGO/VIRGO, two groups
embarked on the derivation of the equations of motion at the third
post-Newtonian (3PN) level. One group used the Arnowitt-Deser-Misner
(ADM) Hamiltonian formalism of general
relativity~\cite{JaraS98,JaraS99,DJSpoinc,DJSequiv} and worked in a
corresponding ADM-type coordinate system. Another group used a direct
post-Newtonian iteration of the equations of motion in harmonic
coordinates~\cite{BF00,BFeom,BFreg,BFregM,ABF01,BI03CM}.  The end
results of these two approaches have been proved to be physically
equivalent~\cite{DJSequiv,ABF01}. However, both approaches, even after
exploiting all symmetries and pushing their
Hadamard-regularization-based methods to the maximum, left
undetermined one dimensionless parameter at the 3PN order. The
appearance of this unknown parameter was related with the choice of
the regularization method used to cure the self-field divergencies of
point particles. Both lines of works regularized the self-field
divergencies by some version of the Hadamard regularization
method. Finally, the completion of the equations of motion at the 3PN
order was made possible thanks to the powerful \textit{dimensional}
self-field regularization, which could fix up uniquely the value of
the ambiguity parameter~\cite{DJSdim,BDE04}.  This result is also in
complete agreement with the recent finding of~\cite{itoh1,itoh2}, who
derived the 3PN equations of motion in harmonic gauge using a
``surface-integral'' approach without use of self-field
regularization. Finally, the 3.5PN terms, which constitute a 1PN
relative modification of the radiation reaction force (and are
relatively easier to derive), have been added in
Refs.~\cite{IW93,IW95,JaraS97,PW02,KFS03,NB05}.

In Section~\ref{secII} of the present paper we give the final result for
the equations of motion of compact binaries at the 3.5PN order. The
equations are presented in ready to use quasi-Newtonian form, in the
reference frame associated with the center of mass position. In
Section~\ref{secIII} we discuss the Lagrangian and Hamiltonian
formulations of the conservative part of the equations, obtained by
neglecting the radiation reaction terms occuring at the 2.5PN and 3.5PN
orders. Finally, in Section~\ref{secIV}, we investigate, following
Ref.~\cite{BI03CM}, the question of the stability of circular orbits,
against linear gravitational perturbations, up to the 3PN order.

\section{Equations of motion in the center-of-mass frame}\label{secII}
In the present paper we employ the so-called harmonic-coordinates
approach~\cite{BF00,BFeom,ABF01,BI03CM} which derived the 3PN binary's
equations of motion in harmonic coordinates. From these equations,
obtained at first in a general frame, we translate the origin of
coordinates to the binary's center-of-mass by imposing that the
binary's center-of-mass vector position is ${\rm G}_i = 0$. The
center-of-mass vector is nothing but the conserved integral of the
motion that is associated, {\it via} the Noether theorem, with the
\textit{boost symmetry} of the Lagrangian from which the 3PN equations
of motion are derived~\cite{ABF01,BI03CM}. The condition ${\rm G}_i =
0$ results in the 3PN-accurate relationship between the individual
positions of the particles $y_1^i$ and $y_2^i$ in the center-of-mass
frame, and the relative position $x^i= y_{1}^i-y_{2}^i$ and velocity
$v^i= v_{1}^i-v_{2}^i = d x^i/dt$. The center-of-mass equations of
motion are then obtained by replacing the individual positions and
velocities by their center-of-mass expressions, given in terms of
$x^i$ and velocity $v^i$, applying as usual the order-reduction of all
accelerations where necessary. Order reduction means that any
acceleration (or derivative of acceleration) in a sub-dominant
post-Newtonian term is to be replaced by its explicit expression given
as a function of the positions and velocities as deduced from the
lower-order equations of motion themselves.

We shall denote the orbital separation by $r=\vert{\bf x}\vert$, and
pose ${\bf n}={\bf x}/r$ and $\dot{r}={\bf n}\cdot{\bf v}$. The mass
parameters are the total mass $m=m_1+m_2$, the mass difference $\delta
m=m_1-m_2$, the reduced mass $\mu=m_1m_2/m$, and the very useful
symmetric mass ratio $\nu=\mu/m$, which is such that $0<\nu\leq 1/4$,
with $\nu=1/4$ in the case of equal masses, and $\nu\to 0$ in the
``test-mass'' limit for one of the bodies. We write the relative
acceleration in the center-of-mass frame in the form (we pose $G=1$)
\begin{equation}\label{eom}
\frac{d v^i}{dt}=-\frac{m}{r^2}\Big[(1+\mathcal{A})\,n^i +
\mathcal{B}\,v^i \Big]+ \mathcal{O}\left( \frac{1}{c^8} \right)\;,
\end{equation}
where the first term represents the famous Newtonian approximation, and
where the post-Newtonian remainder term $\mathcal{O}( c^{-8})$ indicates
the level of accuracy of the expression which is here 3.5PN order. We
find~\cite{BI03CM,NB05} that the coefficients $\mathcal{A}$ and
$\mathcal{B}$ are
\allowdisplaybreaks{\begin{eqnarray}
\mathcal{A}&=& \frac{1}{c^2}\left\{-\frac{3\,\dot{r}^2\,\nu}{2} + v^2 +
3\,\nu\,v^2-\frac{m}{r}\left(4 +2\,\nu \right)\right\}\nonumber\\ &+&
\frac{1}{c^4}\left\{\frac{15\,\dot{r}^4\,\nu}{8} -
\frac{45\,\dot{r}^4\,\nu^2}{8} - \frac{9\,\dot{r}^2\,\nu\,v^2}{2} +
6\,\dot{r}^2\,\nu^2\,v^2 + 3\,\nu\,v^4 - 4\,\nu^2\,v^4\right.
\nonumber\\ &&\qquad + \left.\frac{m}{r}\left( -2\,\dot{r}^2 -
25\,\dot{r}^2\,\nu - 2\,\dot{r}^2\,\nu^2 - \frac{13\,\nu\,v^2}{2} +
2\,\nu^2\,v^2 \right)\right. \nonumber\\ &&\qquad +\left.\frac{m^2}{r^2}\,\left( 9 + \frac{87\,\nu}{4}
\right)\right\}\nonumber\\&+&\frac{1}{c^5}\left\{-
\frac{24\,\dot{r}\,\nu\,v^2}{5}\frac{m}{r}-\frac{136\,\dot{r}\,\nu}{15} \frac{m^2}{r^2}\right\}\nonumber\\ &+&
\frac{1}{c^6}\left\{-\frac{35\,\dot{r}^6\,\nu}{16} +
\frac{175\,\dot{r}^6\,\nu^2}{16} -
\frac{175\,\dot{r}^6\,\nu^3}{16}+\frac{15\,\dot{r}^4\,\nu\,v^2}{2}
\right.\nonumber\\&&\qquad - \left. \frac{135\,\dot{r}^4\,\nu^2\,v^2}{4}
+ \frac{255\,\dot{r}^4\,\nu^3\,v^2}{8} -
\frac{15\,\dot{r}^2\,\nu\,v^4}{2} + \frac{237\,\dot{r}^2\,\nu^2\,v^4}{8}
\right.\nonumber\\ &&\qquad -\left. \frac{45\,\dot{r}^2\,\nu^3\,v^4}{2}
+ \frac{11\,\nu\,v^6}{4} - \frac{49\,\nu^2\,v^6}{4} + 13\,\nu^3\,v^6
\right.\nonumber\\ &&\qquad + \left.\frac{m}{r}\left(
79\,\dot{r}^4\,\nu - \frac{69\,\dot{r}^4\,\nu^2}{2} -
30\,\dot{r}^4\,\nu^3 - 121\,\dot{r}^2\,\nu\,v^2 +
16\,\dot{r}^2\,\nu^2\,v^2 \right.\right.\nonumber\\&&\qquad\qquad\quad~
+\left.\left. 20\,\dot{r}^2\,\nu^3\,v^2+\frac{75\,\nu\,v^4}{4} +
8\,\nu^2\,v^4 - 10\,\nu^3\,v^4 \right)\right.\nonumber\\ &&\qquad +
\left. \frac{m^2}{r^2}\,\left( \dot{r}^2 +
\frac{32573\,\dot{r}^2\,\nu}{168} + \frac{11\,\dot{r}^2\,\nu^2}{8} -
7\,\dot{r}^2\,\nu^3 + \frac{615\,\dot{r}^2\,\nu\,\pi^2}{64} -
\frac{26987\,\nu\,v^2}{840}
\right.\right.\nonumber\\&&\qquad\qquad\quad~ +\left.\left. \nu^3\,v^2 -
\frac{123\,\nu\,\pi^2\,v^2}{64} - 110\,\dot{r}^2\,\nu\,\ln
\Big(\frac{r}{r'_0}\Big) + 22\,\nu\,v^2\,\ln \Big(\frac{r}{r'_0}\Big)
\right)\right.\nonumber\\&&\qquad +\left.\frac{m^3}{r^3}\left( -16 -
\frac{437\,\nu}{4} - \frac{71\,\nu^2}{2} + \frac{41\,\nu\,{\pi }^2}{16}
\right)\right\}\nonumber\\ &+& \frac{1}{c^7}\left\{ \frac{m}{r} \,
\left( \frac{366}{35}\,\nu\,v^4 + 12 \nu^2\,v^4 - 114\,v^2\,\nu
\dot{r}^2 - 12 \nu^2\,v^2 \dot{r}^2 + 112 \nu\,\dot{r}^4\right) \right.
\nonumber\\ & & \qquad\left. + \frac{m^2}{r^2} \, \left(
\frac{692}{35}\,\nu\,v^2 - \frac{724}{15}\,v^2\, \nu^2 +
\frac{294}{5}\,\nu\,\dot{r}^2 + \frac{376}{5} \nu^2\,\dot{r}^2
\right)\right. \nonumber\\ & & \qquad\left. + \frac{m^3}{r^3} \,
\left( \frac{3956}{35}\nu + \frac{184}{5} \nu^2 \right)
\right\},\nonumber\\ \mathcal{B}&=&\frac{1}{c^2}\Big\{-4\,\dot{r} +
2\,\dot{r}\,\nu\Big\}
\nonumber\\&+&\frac{1}{c^4}\left\{\frac{9\,\dot{r}^3\,\nu}{2} +
3\,\dot{r}^3\,\nu^2 -\frac{15\,\dot{r}\,\nu\,v^2}{2} -
2\,\dot{r}\,\nu^2\,v^2\right.\nonumber\\ &&\qquad + \left.\frac{m}{r}\left( 2\,\dot{r} + \frac{41\,\dot{r}\,\nu}{2} + 4\,\dot{r}\,\nu^2
\right)\right\}\nonumber\\&+&\frac{1}{c^5}\left\{
\frac{8\,\nu\,v^2}{5}\frac{m}{r}+\frac{24\,\nu}{5} \frac{m^2}{r^2}\right\}\nonumber\\ &+&
\frac{1}{c^6}\left\{-\frac{45\,\dot{r}^5\,\nu}{8} + 15\,\dot{r}^5\,\nu^2
+ \frac{15\,\dot{r}^5\,\nu^3}{4} + 12\,\dot{r}^3\,\nu\,v^2
\right.\nonumber\\&&\qquad -\left. \frac{111\,\dot{r}^3\,\nu^2\,v^2}{4}
-12\,\dot{r}^3\,\nu^3\,v^2 -\frac{65\,\dot{r}\,\nu\,v^4}{8} +
19\,\dot{r}\,\nu^2\,v^4 + 6\,\dot{r}\,\nu^3\,v^4
\right.\nonumber\\&&\qquad\left. + \frac{m}{r}\left(
\frac{329\,\dot{r}^3\,\nu}{6} + \frac{59\,\dot{r}^3\,\nu^2}{2} +
18\,\dot{r}^3\,\nu^3 - 15\,\dot{r}\,\nu\,v^2 - 27\,\dot{r}\,\nu^2\,v^2 -
10\,\dot{r}\,\nu^3\,v^2 \right) \right.\nonumber\\&&\qquad
+\left.\frac{m^2}{r^2}\,\left( -4\,\dot{r} -
\frac{18169\,\dot{r}\,\nu}{840} + 25\,\dot{r}\,\nu^2 + 8\,\dot{r}\,\nu^3
- \frac{123\,\dot{r}\,\nu\,\pi^2}{32} + 44\,\dot{r}\,\nu\,\ln
\Big(\frac{r}{r'_0}\Big) \right)\right\}\nonumber\\ &+&
\frac{1}{c^7}\left\{ \frac{m}{r} \, \left( - \frac{626}{35}\,\nu\,v^4
- \frac{12}{5} \nu^2\,v^4 + \frac{678}{5}\,\nu\,v^2 \dot{r}^2 +
\frac{12}{5} \nu^2\,v^2 \dot{r}^2 - 120 \nu\,\dot{r}^4\right)
\right.\nonumber\\ & & \qquad\left. + \frac{m^2}{r^2} \, \left(
\frac{164}{21}\,\nu\,v^2 + \frac{148}{5} \nu^2\,v^2 -
\frac{82}{3}\,\nu\,\dot{r}^2 - \frac{848}{15} \nu^2\,\dot{r}^2 \right)
\right.\nonumber\\ & & \qquad\left. + \frac{m^3}{r^3} \, \left( -
\frac{1060}{21}\nu - \frac{104}{5}\nu^2 \right)\right\}.\label{ABcoeff}
\end{eqnarray}}\noindent
The 3.5PN equations of motion play a crucial role when deriving the
high-accuracy templates which will be used for analysing (hopefully in a
near future) the gravitational wave signals from compact binary inspiral
in the data analysis of the LIGO and VIRGO detectors.

At the 3PN order we find some logarithmic terms, depending on some
arbitrary constant $r'_0$. The presence of these logarithms reflects in
fact the use of a specific harmonic coordinate system. It is indeed
known that the logarithms at the 3PN order in Eqs.~(\ref{ABcoeff}),
together with the constant $r'_0$ therein, can be removed by applying a
gauge transformation. This shows that there is no physics associated
with them, and that these logarithms and the constant $r'_0$ will never
appear in any physical result derived from these equations, because the
physical results must be gauge invariant ($r'_0$ is sometimes referred
to as a ``gauge constant''). The gauge transformation at 3PN order whose
effect is to remove the logarithms is given in~\cite{BFeom}. Notice that
after applying this gauge transformation we are still within the class
of harmonic coordinates. The resulting modification of the equations of
motion affects only the coefficients of the 3PN order in
Eqs.~(\ref{ABcoeff}); let us denote them by $\mathcal{A}_{\rm 3PN}$ and
$\mathcal{B}_{\rm 3PN}$. The new values of these coefficients, say
$\mathcal{A}'_{\rm 3PN}$ and $\mathcal{B}'_{\rm 3PN}$, obtained after
removal of the logarithms by the latter harmonic gauge transformation,
are then~\cite{MW03}
\allowdisplaybreaks{\begin{eqnarray} \mathcal{A}'_{\rm 3PN}&=&
\frac{1}{c^6}\left\{-\frac{35\,\dot{r}^6\,\nu}{16} +
\frac{175\,\dot{r}^6\,\nu^2}{16} -
\frac{175\,\dot{r}^6\,\nu^3}{16}+\frac{15\,\dot{r}^4\,\nu\,v^2}{2}
\right.\nonumber\\&&\qquad - \left. \frac{135\,\dot{r}^4\,\nu^2\,v^2}{4}
+ \frac{255\,\dot{r}^4\,\nu^3\,v^2}{8} -
\frac{15\,\dot{r}^2\,\nu\,v^4}{2} + \frac{237\,\dot{r}^2\,\nu^2\,v^4}{8}
\right.\nonumber\\ &&\qquad -\left. \frac{45\,\dot{r}^2\,\nu^3\,v^4}{2}
+ \frac{11\,\nu\,v^6}{4} - \frac{49\,\nu^2\,v^6}{4} + 13\,\nu^3\,v^6
\right.\nonumber\\ &&\qquad + \left.\frac{m}{r}\left(
79\,\dot{r}^4\,\nu - \frac{69\,\dot{r}^4\,\nu^2}{2} -
30\,\dot{r}^4\,\nu^3 - 121\,\dot{r}^2\,\nu\,v^2 +
16\,\dot{r}^2\,\nu^2\,v^2 \right.\right.\nonumber\\&&\qquad\qquad\quad~
+\left.\left. 20\,\dot{r}^2\,\nu^3\,v^2+\frac{75\,\nu\,v^4}{4} +
8\,\nu^2\,v^4 - 10\,\nu^3\,v^4 \right)\right.\nonumber\\ &&\qquad +
\left. \frac{m^2}{r^2}\,\left( \dot{r}^2 +
\frac{22717\,\dot{r}^2\,\nu}{168} + \frac{11\,\dot{r}^2\,\nu^2}{8} -
7\,\dot{r}^2\,\nu^3 + \frac{615\,\dot{r}^2\,\nu\,\pi^2}{64}
\right.\right.\nonumber\\&&\qquad\qquad\quad~ \left.\left. -
\frac{20827\,\nu\,v^2}{840} + \nu^3\,v^2 -
\frac{123\,\nu\,\pi^2\,v^2}{64} \right)\right.\nonumber\\&&\qquad
+\left.\frac{m^3}{r^3}\left( -16 - \frac{1399\,\nu}{12} -
\frac{71\,\nu^2}{2} + \frac{41\,\nu\,{\pi }^2}{16}
\right)\right\},\nonumber\\ \mathcal{B}'_{\rm 3PN}&=&
\frac{1}{c^6}\left\{-\frac{45\,\dot{r}^5\,\nu}{8} + 15\,\dot{r}^5\,\nu^2
+ \frac{15\,\dot{r}^5\,\nu^3}{4} + 12\,\dot{r}^3\,\nu\,v^2
\right.\nonumber\\&&\qquad -\left. \frac{111\,\dot{r}^3\,\nu^2\,v^2}{4}
-12\,\dot{r}^3\,\nu^3\,v^2 -\frac{65\,\dot{r}\,\nu\,v^4}{8} +
19\,\dot{r}\,\nu^2\,v^4 + 6\,\dot{r}\,\nu^3\,v^4
\right.\nonumber\\&&\qquad\left. + \frac{m}{r}\left(
\frac{329\,\dot{r}^3\,\nu}{6} + \frac{59\,\dot{r}^3\,\nu^2}{2} +
18\,\dot{r}^3\,\nu^3 - 15\,\dot{r}\,\nu\,v^2 - 27\,\dot{r}\,\nu^2\,v^2 -
10\,\dot{r}\,\nu^3\,v^2 \right) \right.\nonumber\\&&\qquad
+\left.\frac{m^2}{r^2}\,\left( -4\,\dot{r} -
\frac{5849\,\dot{r}\,\nu}{840} + 25\,\dot{r}\,\nu^2 + 8\,\dot{r}\,\nu^3
- \frac{123\,\dot{r}\,\nu\,\pi^2}{32}
\right)\right\}.\label{ABcoeffprime}
\end{eqnarray}}\noindent
These gauge-transformed coefficients are useful because they do not
yield the usual complications associated with logarithms. However, they
must be handled with care in applications such as~\cite{MW03}, since one
must ensure that all other quantities in the problem (energy, angular
momentum, gravitational-wave fluxes {\it etc.}) are defined in the same
specific harmonic gauge avoiding logarithms. In the following we shall
no longer use the coordinate system leading to
Eqs.~(\ref{ABcoeffprime}). Notably the expression we shall derive below
for the Lagrangian will be valid in the ``standard'' harmonic coordinate
system in which the equations of motion are given by~(\ref{eom})
with~(\ref{ABcoeff}).

\section{Lagrangian and Hamiltonian formulations}\label{secIII}
The Lagrangian for the relative center-of-mass motion is obtained from
the 3PN center-of-mass equations of motion (\ref{eom})--(\ref{ABcoeff})
in which one ignores the radiation-reaction terms at the 2.5PN and 3.5PN
orders. We are indeed interested in the conservative part of the
equations of motion, excluding the terms associated with gravitational
radiation; only the conservative part is deducible from a Lagrangian. It
is known that the Lagrangian in harmonic coordinates will necessarily be
a generalized one (from the 2PN order), {\it i.e.} depending not only on
the positions and velocities of the particles, but also on their
\textit{accelerations}~\cite{DD81b,DS85}. It is also
known~\cite{DS85,ABF01} that one can always restrict ourselves to
a Lagrangian that is linear in the accelerations.

The conservative part of the center-of-mass equations of motion
(\ref{eom})--(\ref{ABcoeff}) then take the form (after systematic
order-reduction of the accelerations) of the generalized Lagrange
equations
\begin{equation}\label{Lequ}
\frac{\partial L}{\partial x^i} -\frac{d}{dt}\left(\frac{\partial
  L}{\partial v^i}\right) +\frac{d^2}{dt^2}\left(\frac{\partial
  L}{\partial a^i}\right)=\mathcal{O}\left( \frac{1}{c^8} \right)\;,
\end{equation}
where $L[x^i, v^i, a^i]$ denotes the generalized center-of-mass
Lagrangian --- which is linear in the accelerations appearing at 2PN and
3PN orders. We recall that there is a large freedom for choosing a
Lagrangian because we can always add to it the total time derivative of
an arbitrary function. As a matter of convenience, we choose below a
particular center-of-mass Lagrangian that is ``close'' (in the sense
that many coefficients are identical) to some ``fictitious'' Lagrangian
obtained from the general-frame one given in Ref.~\cite{ABF01} by the
mere Newtonian center-of-mass replacements $y_1^i\rightarrow
\frac{m_2}{m} x^i$, $y_2^i\rightarrow -\frac{m_1}{m} x^i$. We point out
that such a fictitious Lagrangian is {\it not} the correct Lagrangian
for describing the center-of-mass relative motion. Indeed, the actual
relations connecting the center-of-mass variables $y_1^i$ and $y_2^i$ to
the relative position $x^i$ and velocity $v^i$, involve many
post-Newtonian corrections, so the actual center-of-mass Lagrangian must
contain some extra terms in addition to those of the latter fictitious
one. However, we find that these extra terms arise only from the 2PN
order. Our result (when divided by the reduced mass $\mu=m\nu$) is then
\allowdisplaybreaks{\begin{eqnarray}\label{Lharm} \frac{L}{\mu}&=&
\frac{v^2}{2} + \frac{m}{r} \nonumber\\ &&
+\frac{1}{c^2}\bigg\{\frac{v^4}{8} - \frac{3\,\nu\,v^4}{8} +
\frac{m}{r}\,\left( \frac{\dot{r}^2\,\nu}{2} + \frac{3\,v^2}{2} +
\frac{\nu\,v^2}{2} \right)-\frac{m^2}{2\,r^2} \bigg\}\nonumber\\ &&
+\frac{1}{c^4}\bigg\{ \frac{v^6}{16} - \frac{7\,\nu\,v^6}{16} +
\frac{13\,\nu^2\,v^6}{16} \nonumber\\ &&\qquad~ + \frac{m}{r}\,\left(
\frac{3\,\dot{r}^4\,\nu^2}{8} - \frac{\dot{r}^2\,a_n\,\nu\,r}{8} +
\frac{\dot{r}^2\,\nu\,v^2}{4} - \frac{5\,\dot{r}^2\,\nu^2\,v^2}{4} +
\frac{7\,a_n\,\nu\,r\,v^2}{8} \right.\nonumber\\ &&\qquad\qquad\quad~ +
\left.\frac{7\,v^4}{8} - \frac{5\,\nu\,v^4}{4} - \frac{9\,\nu^2\,v^4}{8}
- \frac{7\,\dot{r}\,\nu\,r\,a_v}{4} \right) \nonumber\\ &&\qquad~
+\frac{m^2}{r^2}\,\left( \frac{\dot{r}^2}{2} +
\frac{41\,\dot{r}^2\,\nu}{8} + \frac{3\,\dot{r}^2\,\nu^2}{2} +
\frac{7\,v^2}{4} - \frac{27\,\nu\,v^2}{8} + \frac{\nu^2\,v^2}{2} \right)
\nonumber\\ &&\qquad~ +\frac{m^3}{r^3}\,\left( \frac{1}{2} +
\frac{15\,\nu}{4} \right)\bigg\}\nonumber\\ &&
+\frac{1}{c^6}\bigg\{\frac{5\,v^8}{128} - \frac{59\,\nu\,v^8}{128} +
\frac{119\,\nu^2\,v^8}{64} - \frac{323\,\nu^3\,v^8}{128} \nonumber\\
&&\qquad~ + \frac{m}{r}\,\left( \frac{5\,\dot{r}^6\,\nu^3}{16} +
\frac{\dot{r}^4\,a_n\,\nu\,r}{16} -
\frac{5\,\dot{r}^4\,a_n\,\nu^2\,r}{16} -
\frac{3\,\dot{r}^4\,\nu\,v^2}{16} \right.\nonumber\\
&&\qquad\qquad\quad~\left.+ \frac{7\,\dot{r}^4\,\nu^2\,v^2}{4} -
\frac{33\,\dot{r}^4\,\nu^3\,v^2}{16} -
\frac{3\,\dot{r}^2\,a_n\,\nu\,r\,v^2}{16} -
\frac{\dot{r}^2\,a_n\,\nu^2\,r\,v^2}{16} \right.\nonumber\\
&&\qquad\qquad\quad~\left.+ \frac{5\,\dot{r}^2\,\nu\,v^4}{8} -
3\,\dot{r}^2\,\nu^2\,v^4 +\frac{75\,\dot{r}^2\,\nu^3\,v^4}{16} +
\frac{7\,a_n\,\nu\,r\,v^4}{8} \right.\nonumber\\
&&\qquad\qquad\quad~\left.- \frac{7\,a_n\,\nu^2\,r\,v^4}{2} +
\frac{11\,v^6}{16} - \frac{55\,\nu\,v^6}{16} + \frac{5\,\nu^2\,v^6}{2}
\right.\nonumber\\ &&\qquad\qquad\quad~ +\left.
\frac{65\,\nu^3\,v^6}{16} + \frac{5\,\dot{r}^3\,\nu\,r\,a_v}{12} -
\frac{13\,\dot{r}^3\,\nu^2\,r\,a_v}{8} \right.\nonumber\\
&&\qquad\qquad\quad~\left.- \frac{37\,\dot{r}\,\nu\,r\,v^2\,a_v}{8} +
\frac{35\,\dot{r}\,\nu^2\,r\,v^2\,a_v}{4} \right) \nonumber\\ &&\qquad~
+ \frac{m^2}{r^2}\,\left( -\frac{109\,\dot{r}^4\,\nu}{144} -
\frac{259\,\dot{r}^4\,\nu^2}{36} + 2\,\dot{r}^4\,\nu^3 -
\frac{17\,\dot{r}^2\,a_n\,\nu\,r}{6} \right.\nonumber\\
&&\qquad\qquad\quad~ +\left. \frac{97\,\dot{r}^2\,a_n\,\nu^2\,r}{12}
+\frac{\dot{r}^2\,v^2}{4} - \frac{41\,\dot{r}^2\,\nu\,v^2}{6} -
\frac{2287\,\dot{r}^2\,\nu^2\,v^2}{48} \right.\nonumber\\
&&\qquad\qquad\quad~ -\left. \frac{27\,\dot{r}^2\,\nu^3\,v^2}{4} +
\frac{203\,a_n\,\nu\,r\,v^2}{12} + \frac{149\,a_n\,\nu^2\,r\,v^2}{6}
\right.\nonumber\\ &&\qquad\qquad\quad~ +\left. \frac{45\,v^4}{16} +
\frac{53\,\nu\,v^4}{24} + \frac{617\,\nu^2\,v^4}{24} -
\frac{9\,\nu^3\,v^4}{4} \right.\nonumber\\ &&\qquad\qquad\quad~ -\left.
\frac{235\,\dot{r}\,\nu\,r\,a_v}{24} +
\frac{235\,\dot{r}\,\nu^2\,r\,a_v}{6} \right) \nonumber\\ &&\qquad~ +
\frac{m^3}{r^3}\,\left( \frac{3\,\dot{r}^2}{2} -
\frac{12041\,\dot{r}^2\,\nu}{420} + \frac{37\,\dot{r}^2\,\nu^2}{4} +
\frac{7\,\dot{r}^2\,\nu^3}{2} - \frac{123\,\dot{r}^2\,\nu\,{\pi }^2}{64}
\right.\nonumber\\ &&\qquad\qquad\quad~ +\left. \frac{5\,v^2}{4} +
\frac{387\,\nu\,v^2}{70} - \frac{7\,\nu^2\,v^2}{4} +
\frac{\nu^3\,v^2}{2} + \frac{41\,\nu\,{\pi }^2\,v^2}{64}
\right.\nonumber\\ &&\qquad\qquad\quad~ \left. + 22\,\dot{r}^2\,\nu\,\ln
\Big(\frac{r}{r'_0}\Big) - \frac{22\,\nu\,v^2}{3}\ln
\Big(\frac{r}{r'_0}\Big) \right)\nonumber\\ &&\qquad~ +
\frac{m^4}{r^4}\,\left( -\frac{3}{8}- \frac{18469\,\nu}{840} +
\frac{22\,\nu}{3}\ln \Big(\frac{r}{r'_0}\Big) \right) \bigg\}\;.
\end{eqnarray}}\noindent
Witness the acceleration terms present at the 2PN and 3PN orders; our
notation is $a_n= {\bf a}\cdot{\bf n}$ and $a_v= {\bf a}\cdot{\bf v}$
for the scalar products between $a^i=dv^i/dt$ and the direction $n^i$
and velocity $v^i$. We recall here that it is in general forbidden to
order-reduce the accelerations in a Lagrangian.

We next consider the problem of the Hamiltonian associated with (the
conservative part of) the equations of
motion~(\ref{eom})--(\ref{ABcoeff}). This problem is not straightforward
in harmonic coordinates because of the presence of accelerations at 2PN
and 3PN orders in the Lagrangian~(\ref{Lharm}). To proceed, the best is
again to change coordinates, and transform the harmonic coordinate
system into a new system which avoids the appearance of the
accelerations terms, and also, as it will turn out, of the logarithms at
the 3PN order. This new coordinate change will thus contain a piece
which is identical to the one we used to remove the logarithms of the
equations of motion in Section~\ref{secII}. However, this new coordinate
system is not harmonic; it was introduced long ago by Arnowitt, Deser \&
Misner in their study of the Hamiltonian formulation of general
relativity and is called the ADM coordinate system. In Ref.~\cite{ABF01}
the ``contact'' transformation between the particles' variables in
harmonic coordinates and those in ADM coordinates was determined. By
contact transformation we mean the relation between the particles'
trajectories $y_A^i(t)$ in one coordinate system, and the corresponding
trajectories $Y_A^i(t)$ in another, say $\delta
y_A^i(t)=Y_A^i(t)-y_A^i(t)$. Notice that in the contact transformation,
the time $t$ is to be viewed as a ``dummy'' variable. The contact
transformation is not a coordinate transformation between the spatial
vectors in both coordinates, but takes also into account the fact that
the time coordinate changes as well; {\it i.e.} $\delta
y_A^i=\xi^i(y_A)-\xi^0(y_A)v_A^i/c$, where $\xi^\mu(y_A)$ denotes the
four-dimensional change between the coordinates, when evaluated at the
position $y_A=[t,y_A^i(t)]$.

There is a unique contact transformation such that the 3PN
harmonic-coordinates Lagrangian~(\ref{Lharm}) is changed into another
Lagrangian whose Legendre transform coincides with the 3PN
ADM-coordinates Hamiltonian derived in~\cite{JaraS98}
(see~\cite{ABF01,BI03CM} for details). In a first stage this yields
the expression for the ADM-coordinates Lagrangian, in which we use
names appropriate to the ADM variables $X^i=x^i+\delta x^i$, which
means the separation distance $R$, the relative square velocity $V^2$,
and the radial velocity $\dot{R}={\bf N}\cdot{\bf V}$. This is an
ordinary Lagrangian, depending only on the positions and velocities
and without accelerations, $L^{\rm ADM}[X^i, V^i]$, and as we said
which is free of logarithms at the 3PN order. Its explicit expression
is
\allowdisplaybreaks{\begin{eqnarray}\label{LADM} \frac{L^{\rm ADM}}{\mu}
&=&\frac{m}{R}+ \frac{V^2}{2} \nonumber\\ &&
+\frac{1}{c^2}\bigg\{\frac{V^4}{8} - \frac{3\,\nu\,V^4}{8} +
\frac{m}{R}\,\left( \frac{\nu\,\dot{R}^2}{2} + \frac{3\,V^2}{2} +
\frac{\nu\,V^2}{2} \right)-\frac{m^2}{2R^2} \bigg\}\nonumber\\ &&
+\frac{1}{c^4}\bigg\{\frac{V^6}{16} - \frac{7\,\nu\,V^6}{16} +
\frac{13\,\nu^2\,V^6}{16} \nonumber\\&&\qquad~+ \frac{m}{R}\,\left(
\frac{3\,\nu^2\,\dot{R}^4}{8} + \frac{\nu\,\dot{R}^2\,V^2}{2} -
\frac{5\,\nu^2\,\dot{R}^2\,V^2}{4} + \frac{7\,V^4}{8} -
\frac{3\,\nu\,V^4}{2} - \frac{9\,\nu^2\,V^4}{8}
\right)\nonumber\\&&\qquad~ + \frac{m^2}{R^2}\,\left(
\frac{3\,\nu\,\dot{R}^2}{2} + \frac{3\,\nu^2\,\dot{R}^2}{2} + 2\,V^2 -
\nu\,V^2 + \frac{\nu^2\,V^2}{2} \right) \nonumber\\&&\qquad~
+\frac{m^3}{R^3}\,\left( \frac{1}{4} + \frac{3\,\nu}{4}
\right)\bigg\}\nonumber\\ && +\frac{1}{c^6}\bigg\{ \frac{5\,V^8}{128} -
\frac{59\,\nu\,V^8}{128} + \frac{119\,\nu^2\,V^8}{64} -
\frac{323\,\nu^3\,V^8}{128}\nonumber\\&&\qquad~+\frac{m}{R}\,\left(
\frac{5\,\nu^3\,\dot{R}^6}{16} + \frac{9\,\nu^2\,\dot{R}^4\,V^2}{16} -
\frac{33\,\nu^3\,\dot{R}^4\,V^2}{16} + \frac{\nu\,\dot{R}^2\,V^4}{2} -
3\,\nu^2\,\dot{R}^2\,V^4 \right.\nonumber\\&&\qquad\qquad\quad~\left. +
\frac{75\,\nu^3\,\dot{R}^2\,V^4}{16} + \frac{11\,V^6}{16} -
\frac{7\,\nu\,V^6}{2} + \frac{59\,\nu^2\,V^6}{16} +
\frac{65\,\nu^3\,V^6}{16} \right)\nonumber\\&&\qquad~
+\frac{m^2}{R^2}\,\left( - \frac{5\,\nu\,\dot{R}^4}{12} +
\frac{17\,\nu^2\,\dot{R}^4}{12} + 2\,\nu^3\,\dot{R}^4 +
\frac{39\,\nu\,\dot{R}^2\,V^2}{16} - \frac{29\,\nu^2\,\dot{R}^2\,V^2}{8}
\right.\nonumber\\&&\qquad\qquad\quad~\left. -
\frac{27\,\nu^3\,\dot{R}^2\,V^2}{4} + \frac{47\,V^4}{16} -
\frac{15\,\nu\,V^4}{4} - \frac{25\,\nu^2\,V^4}{16} -
\frac{9\,\nu^3\,V^4}{4} \right) \nonumber\\&&\qquad~
+\frac{m^3}{R^3}\,\left( \frac{77\,\nu\,\dot{R}^2}{16} +
\frac{5\,\nu^2\,\dot{R}^2}{4} + \frac{7\,\nu^3\,\dot{R}^2}{2} +
\frac{3\,\nu\,\dot{R}^2\,\pi^2}{64} + \frac{13\,V^2}{8}
\right.\nonumber\\&&\qquad\qquad\quad~\left. - \frac{409\,\nu\,V^2}{48}
- \frac{5\,\nu^2\,V^2}{8} + \frac{\nu^3\,V^2}{2} - \frac{\nu\,{\pi
}^2\,V^2}{64} \right) \nonumber\\&&\qquad~ +\frac{m^4}{R^4}\,\left(
-\frac{1}{8} - \frac{109\,\nu}{12} + \frac{21\,\nu\,\pi^2}{32} \right)
\bigg\}\;.
\end{eqnarray}}\noindent
Next we apply the ordinary Legendre transform to obtain the
corresponding Hamiltonian, $H^{\rm ADM}[X^i, P^i]$, which is a function
of $X^i$ and the conjugate momentum $P^i = \partial L^{\rm ADM}/\partial
V^i$. We find
\allowdisplaybreaks{\begin{eqnarray}\label{HADM} \frac{H^{\rm ADM}}{\mu}
&=& \frac{P^2}{2} -\frac{m}{R}\nonumber\\ && +\frac{1}{c^2}\bigg\{-
\frac{P^4}{8} + \frac{3\,\nu\,P^4}{8} + \frac{m}{R}\left( -
\frac{{P_R}^2\,\nu}{2} - \frac{3\,P^2}{2} - \frac{\nu\,P^2}{2}
\right)+\frac{m^2}{2R^2} \bigg\}\nonumber\\ &&
+\frac{1}{c^4}\bigg\{\frac{P^6}{16} - \frac{5\,\nu\,P^6}{16} +
\frac{5\,\nu^2\,P^6}{16} \nonumber\\&&\qquad~+ \frac{m}{R}\left( -
\frac{3\,{P_R}^4\,\nu^2}{8} - \frac{{P_R}^2\,P^2\,\nu^2}{4} +
\frac{5\,P^4}{8} - \frac{5\,\nu\,P^4}{2} - \frac{3\,\nu^2\,P^4}{8}
\right)\nonumber\\&&\qquad~ + \frac{m^2}{R^2}\,\left(
\frac{3\,{P_R}^2\,\nu}{2} + \frac{5\,P^2}{2} + 4\,\nu\,P^2 \right)
\nonumber\\&&\qquad~+\frac{m^3}{R^3}\left( -\frac{1}{4} -
\frac{3\,\nu}{4} \right) \bigg\}\nonumber\\ &&
+\frac{1}{c^6}\bigg\{-\frac{5\,P^8}{128} + \frac{35\,\nu\,P^8}{128} -
\frac{35\,\nu^2\,P^8}{64} + \frac{35\,\nu^3\,P^8}{128}
\nonumber\\&&\qquad~ + \frac{m}{R}\left( -\frac{5\,{P_R}^6\,\nu^3}{16} +
\frac{3\,{P_R}^4\,P^2\,\nu^2}{16} - \frac{3\,{P_R}^4\,P^2\,\nu^3}{16} +
\frac{{P_R}^2\,P^4\,\nu^2}{8} \right.\nonumber\\&&\qquad\qquad\quad~
\left. - \frac{3\,{P_R}^2\,P^4\,\nu^3}{16}-\frac{7\,P^6}{16} +
\frac{21\,\nu\,P^6}{8} - \frac{53\,\nu^2\,P^6}{16} -
\frac{5\,\nu^3\,P^6}{16} \right) \nonumber\\&&\qquad~ +
\frac{m^2}{R^2}\,\left( \frac{5\,{P_R}^4\,\nu}{12} +
\frac{43\,{P_R}^4\,\nu^2}{12} + \frac{17\,{P_R}^2\,P^2\,\nu}{16}
\right.\nonumber\\&&\qquad\qquad\quad~ \left.+
\frac{15\,{P_R}^2\,P^2\,\nu^2}{8} - \frac{27\,P^4}{16} +
\frac{17\,\nu\,P^4}{2} + \frac{109\,\nu^2\,P^4}{16} \right)
\nonumber\\&&\qquad~ + \frac{m^3}{R^3}\,\left(
-\frac{85\,{P_R}^2\,\nu}{16} - \frac{7\,{P_R}^2\,\nu^2}{4} -
\frac{25\,P^2}{8} - \frac{335\,\nu\,P^2}{48}
\right.\nonumber\\&&\qquad\qquad\quad~ \left.- \frac{23\,\nu^2\,P^2}{8}
- \frac{3\,{P_R}^2\,\nu\,\pi^2}{64} + \frac{\nu\,P^2\,\pi^2}{64}
\right)\nonumber\\&&\qquad~ + \frac{m^4}{R^4}\, \left( \frac{1}{8} +
\frac{109\,\nu}{12} - \frac{21\,\nu\,\pi^2}{32} \right) \bigg\}\;.
\end{eqnarray}}\noindent
We denote $P^2 = {\bf P}^2$ and $P_R = {\bf N}\cdot{\bf P}$. The
previous result is in perfect agreement with the center-of-mass
Hamiltonian derived in Ref.~\cite{JaraS98}.

\section{Dynamical stability of circular orbits}\label{secIV}
As an application let us investigate the problem of the stability,
against dynamical gravitational perturbations, of circular orbits at
the 3PN order. We want in particular to discuss the existence (or
non-existence) of an innermost stable circular orbit (ISCO) at various
post-Newtonian orders, which would constitute the analogue for two
black holes with comparable masses of the famous orbit
$r_\mathrm{ISCO}=6M/c^2$ in the Schwarzschild metric. Notice that we
are concerned here with the stability of the orbit with respect to
\textit{purely gravitational} perturbations appropriate to the motion
of black holes; however it is known that for neutron stars instead of
black holes the ISCO is determined by the hydrodynamical instability
rather than by the effect of general relativity.

We propose to use two different methods for this problem, one based on
a perturbation at the level of the equations of motion
(\ref{eom})--(\ref{ABcoeff}) in harmonic coordinates, the other one
consisting of perturbing the Hamiltonian equations in ADM coordinates
for the Hamiltonian~(\ref{HADM}). We shall find a criterion for the
stability of orbits and shall present it in the form of an invariant
expression (which is the same in different coordinate systems). We
shall check that our two methods agree on the result.

We deal first with the perturbation of the equations of motion,
following the approach proposed in Section III.A of
Ref.~\cite{KWW93}. We introduce polar coordinates $(r,\varphi)$ in the
orbital plane and pose $u= {\dot r}$ and $\omega= {\dot
\varphi}$. Then Eq.~(\ref{eom}) yields the system of equations
\begin{subequations}\label{36}\begin{eqnarray}
{\dot r} &=& u\;,\label{36a}\\ {\dot u} &=&
-\frac{m}{r^2}\Big[1+\mathcal{A}+\mathcal{B}u\Big]+r\omega^2\;,\label{36b}\\
{\dot \omega} &=& -\omega\left[\frac{m}{r^2}\mathcal{B}+\frac{2
	u}{r}\right]\;,\label{36c}
\end{eqnarray}\end{subequations}
where $\mathcal{A}$ and $\mathcal{B}$ are given by Eqs.~(\ref{ABcoeff})
as functions of $r$, $u$ and $\omega$ (through $v^2=u^2+r^2\omega^2$).
In the case of an orbit which is circular, apart from the adiabatic
inspiral due to the 2.5PN and 3.5PN radiation-reaction effects, we have
$\dot r=\dot u=\dot \omega=0$ hence $u=0$. Eq.~(\ref{36b}) gives thereby
the angular velocity $\omega_0$ of the circular orbit as
\begin{equation}\label{37}
\omega_0^2 = \frac{m}{r_0^3}\big(1+\mathcal{A}_0\big)\;.
\end{equation}
Solving iteratively this relation at the 3PN order using the equations
of motion (\ref{eom})--(\ref{ABcoeff}) we obtain $\omega_0$ as a
function of the circular-orbit radius $r_0$ in harmonic coordinates (the
result agrees with the one of Refs.~\cite{BF00,BFeom}),
\begin{eqnarray}\label{42}
\omega_0^2 =
\frac{m}{r_0^3}\Bigg\{1&+&\frac{m}{r_0\,c^2}\Big(-3+\nu\Big)
+\frac{m^2}{r_0^2\, c^4}\left(6+\frac{41}{4}\nu+\nu^2\right)\nonumber\\
&+&\frac{m^3}{r_0^3\,c^6}\left(-10
+\left[-\frac{75707}{840}+\frac{41}{64}\pi^2+22\ln
\Big(\frac{r_0}{r'_0}\Big)\right]\nu+\frac{19}{2}\nu^2+\nu^3\right) +
\mathcal{O}\left(\frac{1}{c^8}\right)\Bigg\}\;.
\end{eqnarray}
The circular-orbit radius $r_0$ should not be confused with the constant
$r'_0$ entering the logarithm at the 3PN order and which is issued from
Eqs.~(\ref{ABcoeff}).

Now we investigate the equations of linear perturbations around the
circular orbit defined by the constants $r_0$, $u_0=0$, or, rather, if
we were to include the radiation-reaction damping,
$u_0=\mathcal{O}(c^{-5})$, and $\omega_0$. We pose
\begin{subequations}\label{38}\begin{eqnarray}
r &=& r_0 + \delta r\;,\\
u &=& \delta u\;,\\
\omega &=& \omega_0 + \delta \omega\;,
\end{eqnarray}\end{subequations}
where $\delta r$, $\delta u$ and $\delta \omega$ denote some
perturbations of the circular orbit. Then a system of linear equations
follows as
\begin{subequations}\label{39}\begin{eqnarray}
\dot{\delta r} &=& \delta u\;,\\ \dot{\delta u} &=& \alpha_0\, \delta r
+ \beta_0\, \delta \omega\;,\\ \dot{\delta \omega} &=& \gamma_0\, \delta
u\;,
\end{eqnarray}\end{subequations}
where the coefficients, which solely depend on the unperturbed circular
orbit, read~\cite{KWW93}
\begin{subequations}\label{40}\begin{eqnarray}
\alpha_0 &=& 3 \omega_0^2 - \frac{m}{r_0^2}\left(\frac{\partial
\mathcal{A}}{\partial r}\right)_0\;,\\ \beta_0 &=& 2 r_0 \omega_0 -
\frac{m}{r_0^2}\left(\frac{\partial \mathcal{A}}{\partial
\omega}\right)_0\;,\\ \gamma_0 &=& -\omega_0 \left[\frac{2}{r_0} +
\frac{m}{r_0^2}\left(\frac{\partial \mathcal{B}}{\partial
u}\right)_0\right]\;.
\end{eqnarray}\end{subequations}
In obtaining Eqs. (\ref{40}) we use the fact that $\mathcal{A}$ is a
function of the square $u^2$ through $v^2=u^2+r^2\omega^2$, so that
$\partial \mathcal{A}/\partial u$ is proportional to $u$ and thus
vanishes in the unperturbed configuration (because $u=\delta u$). On the
other hand, since the radiation reaction is neglected, $\mathcal{B}$
also is proportional to $u$ [see Eqs.~(\ref{ABcoeff})], so only
$\partial \mathcal{B}/\partial u$ can contribute at the zeroth
perturbative order. Now by examining the fate of perturbations that are
proportional to some $e^{i\sigma t}$, we arrive at the condition for the
frequency $\sigma$ of the perturbation to be real, and hence for stable
circular orbits to exist, as being~\cite{KWW93}
\begin{equation}\label{41}
\hat{C}_0 = -\alpha_0 - \beta_0\, \gamma_0 ~> 0\;.
\end{equation}
Substituting into this $\mathcal{A}$ and $\mathcal{B}$ at the 3PN order we
then arrive at the orbital-stability criterion
\begin{eqnarray}\label{43}
\hat{C}_0 = \frac{m}{r_0^3}\Bigg\{1&+&\frac{m}{r_0\,c^2}\Big(-9+\nu\Big)
+\frac{m^2}{r_0^2\,c^4}\left(30 +\frac{65}{4}\nu+\nu^2\right)\nonumber\\
&+&\frac{m^3}{r_0^3\,c^6}
\left(-70+\left[-\frac{29927}{840}-\frac{451}{64}\pi^2+22\ln
\Big(\frac{r_0}{r'_0}\Big) \right]\nu+\frac{19}{2}\nu^2+\nu^3\right) +
\mathcal{O}\left(\frac{1}{c^8}\right)\Bigg\}\;,
\end{eqnarray}
where we recall that $r_0$ is the radius of the orbit in harmonic
coordinates.

Our second method is to use the Hamiltonian equations based on the 3PN
Hamiltonian in ADM coordinates given by Eq.~(\ref{HADM}). We introduce
the polar coordinates $(R,\Psi)$ in the orbital plane --- we assume that
the orbital plane is equatorial, given by $\Theta=\frac{\pi}{2}$ in the
spherical coordinate system $(R,\Theta,\Psi)$ --- and make the
substitution
\begin{equation}\label{44}
P^2={P_R}^2+\frac{P_\Psi^2}{R^2}\;,
\end{equation}
into the Hamiltonian. This yields a ``reduced'' Hamiltonian that is a
function of $R$, $P_R$ and $P_\Psi$, namely
$\mathcal{H}=\mathcal{H}\big[R,P_R,P_\Psi\big]$, and describes the
motion in polar coordinates in the orbital plane (henceforth we denote
$\mathcal{H}=H^{\rm ADM}/\mu$). The Hamiltonian equations then read
\begin{subequations}\label{45}\begin{eqnarray}
\frac{dR}{dt} &=& \frac{\partial \mathcal{H}}{\partial P_R}\;,\\
\frac{d\Psi}{dt} &=& \frac{\partial \mathcal{H}}{\partial P_\Psi}\;,\\
\frac{dP_R}{dt} &=& -\frac{\partial \mathcal{H}}{\partial R}\;,\\
\frac{dP_\Psi}{dt} &=& 0\;.
\end{eqnarray}\end{subequations}
Evidently the constant $P_\Psi$ is nothing but the conserved
angular-momentum integral. For circular orbits we have $R=R_0$ (a
constant) and $P_R=0$, so
\begin{equation}\label{46}
\frac{\partial \mathcal{H}}{\partial R}\big[R_0,0,P_\Psi^0\big] = 0\;,
\end{equation}
which gives the angular momentum $P_\Psi^0$ of the circular orbit as a
function of $R_0$, and
\begin{equation}\label{47}
\omega_0 = \left(\frac{d\Psi}{dt}\right)_0 = \frac{\partial
\mathcal{H}}{\partial P_\Psi}\big[R_0,0,P_\Psi^0\big]\;,
\end{equation}
which yields the angular frequency of the circular orbit $\omega_0$ ---
the same as in Eq. (\ref{42}) --- in terms of $R_0$,
\begin{eqnarray}
\omega_0^2 =
\frac{m}{R_0^3}\Bigg\{1&+&\frac{m}{R_0\,c^2}\Big(-3+\nu\Big)
+\frac{m^2}{R_0^2\,
c^4}\left(\frac{21}{4}-\frac{5}{8}\nu+\nu^2\right)\nonumber\\
&+&\frac{m^3}{R_0^3\,c^6}\left(-7
+\left[-\frac{2015}{48}+\frac{167}{64}\pi^2
\right]\nu-\frac{31}{8}\nu^2+\nu^3\right) +
\mathcal{O}\left(\frac{1}{c^8}\right)\Bigg\}\;.
\end{eqnarray}
The last equation, which is equivalent to $R={\rm const}=R_0$, {\it
i.e.}
\begin{equation}
\frac{\partial \mathcal{H}}{\partial P_R}\big[R_0,0,P_\Psi^0\big] = 0\;,
\end{equation}
is automatically verified because $\mathcal{H}$ is a quadratic function
of $P_R$ and hence $\partial \mathcal{H}/\partial P_R$ is zero for
circular orbits.

We consider now a perturbation of the circular orbit defined by
\begin{subequations}\label{49}\begin{eqnarray}
P_R &=& \delta P_R\;,\\ P_\Psi &=& P_\Psi^0 + \delta P_\Psi\;,\\ R &=&
R_0 + \delta R\;,\\ \omega &=& \omega_0 + \delta \omega\;.
\end{eqnarray}\end{subequations}
It is easy to verify that the Hamiltonian equations~(\ref{45}), when
worked out at the linearized order, read as
\begin{subequations}\label{50}\begin{eqnarray}
\dot{\delta P_R} &=& -\pi_0\, \delta R - \rho_0\, \delta P_\Psi\;,\\
\dot{\delta P_\Psi} &=& 0\;,\\ \dot{\delta R} &=& \sigma_0\, \delta
P_R\;,\\ \delta \omega &=& \rho_0\, \delta R + \tau_0\, \delta P_\Psi\;,
\end{eqnarray}\end{subequations}
where the coefficients, which depend on the unperturbed orbit, are given
by
\begin{subequations}\label{51}\begin{eqnarray}
\pi_0&=&\frac{\partial^2 \mathcal{H}}{\partial
R^2}\big[R_0,0,P_\Psi^0\big]\;,\\ \rho_0&=&\frac{\partial^2
\mathcal{H}}{\partial R\, \partial P_\Psi}\big[R_0,0,P_\Psi^0\big]\;,\\
\sigma_0&=&\frac{\partial^2 \mathcal{H}}{\partial
{P_R}^2}\big[R_0,0,P_\Psi^0\big]\;,\\ \tau_0&=&\frac{\partial^2
\mathcal{H}}{\partial {P_\Psi}^2}\big[R_0,0,P_\Psi^0\big]\;.
\end{eqnarray}\end{subequations}
By looking to solutions proportional to some $e^{i\sigma t}$ one obtains
some real frequencies, and therefore one finds stable circular orbits,
if and only if
\begin{equation}\label{52}
\hat{C}_0 = \pi_0\, \sigma_0 ~> 0\;.
\end{equation}
Using the Hamiltonian (\ref{HADM}) we readily obtain
\begin{eqnarray}\label{53}
\hat{C}_0 = \frac{m}{R_0^3}\Bigg\{1&+&\frac{m}{R_0\,c^2}(-9+\nu)
+\frac{m^2}{R_0^2\,c^4}\left(\frac{117}{4}
+\frac{43}{8}\nu+\nu^2\right)\nonumber\\
&+&\frac{m^3}{R_0^3\,c^6}\left(-61+\left[\frac{4777}{48}
-\frac{325}{64}\pi^2 \right]\nu-\frac{31}{8}\nu^2+\nu^3\right) +
\mathcal{O}\left(\frac{1}{c^8}\right)\Bigg\}\;.
\end{eqnarray}
This result does not look the same as our previous result (\ref{43}),
but this is simply due to the fact that it depends on the ADM radial
separation $R_0$ instead of the harmonic one $r_0$. Fortunately all the
material needed to connect $R_0$ to $r_0$ with the 3PN accuracy is
known~\cite{BI03CM}. In the case of circular orbits we readily find
\begin{eqnarray}\label{54}
R_0 = r_0\Bigg\{1&+&\frac{m^2}{r_0^2\,
c^4}\left(-\frac{1}{4}-\frac{29}{8}\nu\right) +\frac{m^3}{r_0^3\,
c^6}\left(\left[\frac{3163}{1680} +\frac{21}{32}\pi^2
-\frac{22}{3}\ln\Big(\frac{r_0}{r'_0}\Big)\right]\nu
+\frac{3}{8}\nu^2\right) +
\mathcal{O}\left(\frac{1}{c^8}\right)\Bigg\}\;.
\end{eqnarray}
The difference between $R_0$ and $r_0$ is made out of 2PN and 3PN terms
only. Inserting Eq. (\ref{54}) into Eq. (\ref{53}) and re-expanding to
3PN order we find that indeed our basic stability-criterion function
$\hat{C}_0$ comes out the same with our two methods.

Finally let us give to the function $\hat{C}_0$ an invariant meaning by
expressing it with the help of the orbital frequency $\omega_0$ of the
circular orbit, or, more conveniently, of the frequency-related
parameter
\begin{equation}\label{55}
x_0 = \left(\frac{m\,\omega_0}{c^3}\right)^{2/3}\;.
\end{equation} 
From the inverse of Eq.~(\ref{42}) we readily obtain $r_0$ as a function
of $x_0$. This allows us to write the criterion for stability as $C_0 >
0$, where $C_0=\frac{m^2}{c^6\,x_0^3}\hat{C}_0$ admits the
gauge-invariant form, which will be the same in all coordinate systems,
\begin{equation}\label{57}
C_0 = 1-6\,x_0 + 14\,\nu\,x_0^2 +
\left(\left[\frac{397}{2}-\frac{123}{16}\pi^2\right]\nu-14\nu^2\right)\,x_0^3
+ \mathcal{O}\left(x_0^4\,\right)\;.
\end{equation} 
This form is more interesting than the coordinate-dependent expressions
(\ref{43}) or (\ref{53}), not only because of its invariant form, but
also because as we see the 1PN term yields exactly the Schwarzschild
result that the innermost stable circular orbit or ISCO of a test
particle ({\it i.e.} in the limit $\nu\to 0$) is located at $x_{\rm
ISCO}=1/6$. Thus we find that, at the 1PN order, but for {\it any} mass
ratio $\nu$,
\begin{equation}\label{58}
x_{\rm ISCO}^{\rm 1PN} = \frac{1}{6}\;.
\end{equation}
One could have expected that some deviations of the order of $\nu$
already occur at the 1PN order, but it turns out that only from the 2PN
order does one find the occurence of some non-Schwarzschildian
corrections proportional to $\nu$. At the 2PN order we obtain
\begin{equation}
x_{\rm ISCO}^{\rm 2PN} =
\frac{3}{14\nu}\Bigg(1-\sqrt{1-\frac{14\nu}{9}}~\Bigg)\;.
\end{equation}
For equal masses this gives $x_{\rm ISCO}^{\rm 2PN}\simeq 0.187$. Notice
also that the effect of the finite mass corrections is to increase the
frequency of the ISCO with respect to the Schwarzschild result ({\it
i.e.} to make it more inward), and we find $x_{\rm ISCO}^{\rm
2PN}=\frac{1}{6}\left[1+\frac{7}{18}\nu+\mathcal{O}(\nu^2)\right]$.
Finally, at the 3PN order and for equal masses $\nu=\frac{1}{4}$, we
find that according to our criterion all the circular orbits are stable,
and there is no ISCO. More generally, we find that at the 3PN order all
orbits are stable when the mass ratio is $\nu > \nu_c$ where $\nu_c
\simeq 0.183$.

Note that the above stability criterion $C_0$ gives an innermost stable
circular orbit, when it exists, that is not necessarily the same as ---
and actually differs from --- the innermost circular orbit or ICO, which
is defined by the point at which the center-of-mass binding energy of
the binary for circular orbits reaches its minimum value~\cite{B02ico}.
In this respect the present formalism, which is based on systematic
post-Newtonian expansions (without using post-Newtonian resummation
techniques like Pad\'e approximants~\cite{DIS98}), differs from some
``Schwarzschild-like'' methods such as the effective-one-body
approach~\cite{BuonD98} in which the ICO happens to be also an innermost
stable circular orbit or ISCO.

As a final comment, let us note that the use of a {\it truncated}
post-Newtonian series such as Eq.~(\ref{57}) to determine the ISCO is
{\it a priori} meaningful only if we are able to bound the neglected
error terms. Furthermore, since we are dealing with a stability
criterion, it is not completely clear that the higher-order
post-Newtonian correction terms, even if they are numerically small,
will not change qualitatively the response of the orbit to the dynamical
perturbation. This is maybe a problem, and which cannot be answered
rigorously with the present formalism. However, in the regime of the
ISCO (when it exists), we have seen that $x_0$ is rather small,
$x_0\simeq 0.2$ (this is also approximately the value for the ICO
computed in Ref.~\cite{B02ico}), which indicates that the neglected
terms in the truncated series (\ref{57}) should not contribute very
much, because they involve at least a factor $x_0^4\simeq 0.002$. On the
other hand, we pointed out that in the limit $\nu\to 0$ the criterion
$C_0$ gives back the correct {\it exact} result, $x_{\rm ISCO}^{\nu\to
0} = \frac{1}{6}$. This contrasts with the gauge-dependent power series
(\ref{43}) or (\ref{53}) which give only some approximate results. Based
on these observations, we feel that it is reasonable to expect that the
gauge-invariant stability criterion defined by Eq. (\ref{57}) is
physically meaningful.





\section*{Acknowledgements}
This article is based on the work~\cite{BI03CM} which has been done in
collaboration with Bala R. Iyer.

\bibliography{/home/blanchet/Articles/ListeRef/ListeRef}




\end{document}